\documentclass{ws-ijmpa}

\begin{document}

\markboth{S. A. Mart\'\i nez, R. Montemayor and L. F. Urrutia}
{Perturbative Hamiltonian constraints for higher order theories}

\catchline{}{}{}{}{}
%
%%%%%%%%%%%%%%%%%%%%%%%%%%%%%%%%%%%%%%%%%%%%%%%%%%%%%%%%%%%%%%%%%%%%

\title{Perturbative Hamiltonian constraints for higher order theories.}

\author{S. A. Mart\'\i nez, R. Montemayor}

\address{Instituto Balseiro and CAB, Universidad Nacional de Cuyo and CNEA,
8400 Bariloche, Argentina}

\author{L. F. Urrutia}

\address{Instituto de Ciencias Nucleares, Universidad Nacional Aut{\'o}noma de M{\'e}xico, 
A. Postal 70-543, 04510 M{\'e}xico D.F., M{\'e}xico}

\maketitle

\begin{history}
\received{Day Month Year}
\revised{Day Month Year}
\end{history}

\begin{abstract}
We present an alternative method for constructing a consistent perturbative low energy canonical
formalism for higher order time-derivative theories, which consists in appliying the standard
Dirac method to the first order version of the higher order Lagrangian, augmented by additional
perturbative Hamiltonian constraints. The method is purely algebraic, provides the dynamical
formulation directly in phase space and can be used in  singular theories without the need of
initially fixing the gauge. We apply it to two paradigmatic examples: the Pais-Uhlenbeck oscillator
and the Bernard-Duncan scalar field with self-interaction. We also compare the results, both at the
classical and quantum level, with the ones corresponding to a direct perturbative construction
applied to the exact higher order theory, after incorporating  the projection to the space of
physical modes. This comparison highligths the soundness of the present formalism.

\keywords{higher order theories; perturbative canonical formalism}
\end{abstract}

\ccode{PACS numbers: 11.25.Hf, 123.1K}

\section{Introduction}	

Higher order time-derivative (HOTD) theories, including nonlocal theories,
have a long history in physics. They continually reappear in new models of
physical interest, particularly in the form of effective Lagrangians that
describe small corrections to well established theories. For example,
nonlocal effective field theories emerge when high energy degrees of freedom
are integrated out\cite{BV}. HOTD terms also appear in higher derivative
gravity\cite{KSS}, nonconmutative field theory\cite{CDS}, models derived
form string theory\cite{EW,EW1,HH}, effective models for meson nucleon
interactions\cite{KMD}, etc. Our concern in this work is the
discussion of HOTD as perturbations of well established standard second
order theories (to be called precursors), instead of considering them as
fundamental ones. Then it is appropriate to view the resulting HOTD
description as an effective field theory, valid within a given energy range,
where a perturbative description is perfectly acceptable, besides of
constituting a natural setting for extracting physical predictions.

Classical HOTD theories introduce more degrees of freedom than
their precursors, which is better reflected in
their Hamiltonian structure. One characteristic of these theories is the
Ostrogradsky instability, the existence of unphysical runaway solutions, not
expandible in powers of the parameter codifying the HOTD contribution. At
the quantum level the changes in the canonical structure produce important
differences, associated to several pathologies. Even when the HOTD terms are
considered as small corrections to their precursors, as it is done
here, their effect is qualitatively significant. We understand here
the number of degrees of freedom as the number of initial conditions that
must be given to fully determine the behavior of the system at any time.
Naturally, this depends on the degree of the time derivative in the equation
of motion, irrespective of the fact that we might, for example, have only
one coordinate $x(t)$.

The most traditional approach for a canonical formalism for HOTD theories
was developed by Ostrogradsky in 1850\cite{Ost} in the context of regular
theories. This construction is equivalent to finding a second order
Lagrangian by introducing the appropriate auxiliary variables in the HOTD
theories via Lagrange multipliers, and by applying the Dirac method\cite{JP}.
The canonical formalism thus obtained highlights the well known problems of
these theories, such as instability, Hamiltonians unbounded from
below and lack of unitarity at the quantum level. The Ostrogradsky method
was extended to HOTD singular theories by Nesterenko\cite{NESTERENKO}.

However, the HOTD terms usually appear in the Lagrangian as small
corrections, labeled by a small parameter $\gamma$. For this
reason, perturbative schemes have been developed which identify the correct
low-energy degrees of freedom contained in the theory and avoid the
high-energy ones that produce all the inconvenience\cite{EW,CHENGHO,JLM}. A
typical recent example of this is the Myers-Pospelov model\cite{MP}, whereby
HOTD operators of dimension five are introduced as small\ corrections to
standard electrodynamics in order to describe possible minute signals of \
Lorentz invariance violation. An analogous situation occurs in Lorentz
violating extensions for the Standard Model\cite{SME}, which generalizes the
previous situation to the standard model of particles plus gravity. HOTD theories
are also present in the analysis of precision tests of electroweak interactions \cite{ALTARELLI}.
Those  research topics are the subject of a considerable number of
experimental observations with ever increasing precision. A very useful toy
model for exploring the different aspects of such perturbative methods is
the Bernard-Duncan field\cite{BD}, a generalization to field theory of the
Pais-Uhlenbeck quantum mechanical model\cite{PU}, a paradigmatic example of
HOTD theories.

There are two main approaches for reducing a higher order Lagrangian to
one containing only the low energy modes. One of them is based on
the use of field transformations containing derivatives\cite%
{WEINBERG,GEORGI,GROSSE,BaruaG}. The other approach is based on a Lagrangian
which contains HOTD up to a certain order. The dynamics can be reduced to a
second order one by introducing perturbative constraints. Within a
Lagrangian perspective, these constraints become naturally generated by
equations of motion. One of the explored possibilities is to eliminate the
higher order derivatives in the Lagrangian using the equations of motion.
Given that in general the equations of motion can not be introduced in the
Lagrangian without distorting the variational principle, this approach can
be applied only to certain special cases to obtain an approximate second
order Lagrangian\cite{SCHAFER,BARKEROCONELL} Another possibility, explored
in \cite{JLM}, is to project the Lagrangian constraints into the phase space
of momenta and coordinates, and consider these projections as Hamiltonian
constraints, restricting the Ostrogradsky Hamiltonian. The key this approach
is that the perturbative Lagrangians constraints are truly projectable and
form a set of second class constraints. But, as has been shown in \cite%
{PONS1}, Lagrangian and Hamiltonian constraints are not directly related,
and in general not projectable in the case of gauge theories. If there is a
gauge symmetry this approach is not directly applicable, because the first
class constraints are not projectable. For this reason it is necessary to
first fix the gauge at the Lagrangian level to proceed with the construction.

Once the role of the Lagrangian constraints in the reduction of the phase
space is appreciated, the emphasis of the problem shifts to finding more
efficient methods of calculating the iterative steps which are required to
obtain the sought approximation to a given order in $\gamma $. One of
these alternatives is proposed in Refs. \refcite{EW,EW1}, where the construction is
performed basically in the coordinates-velocities (CV) space,
avoiding an explicit projection of constraints. The perturbative Lagrangian
constraints are directly implemented on the Noether energy of the
HOTD theory, which become the time evolution generator in the constrained
CV space. To find the canonical structure this constrained energy is
considered as the corresponding Hamiltonian, and the dynamics is written in
terms of generalized brackets among coordinates and velocities. Imposing
that these brackets reproduce the perturbative equations of motion
to the order considered, the corresponding algebra is determined. This step
requires the solution of a set of second order differential equations, which
become very involved beyond the first order in $\gamma$ and whose solution
involves a good amount of guess work. The algebra in the CV space is
subsequently rewritten in terms of canonical coordinates and momentum.
Ref. \refcite{SIMON} includes a very clear review of this work, together with
that concerning the general problem of HOTD systems. There, the need to add
constraints in order to make perturbative sense of HOTD theories is
also emphasized. Constrained HOTD theories become free from the diseases
that plague unconstrained HOTD ones. This has also been previously remarked
in Ref. \refcite{BHABBA}. Finally we mention Ref. \refcite{CHENGHO}, where the use
of iterative solutions of the equation of motion is focused on the obtention
of the final symplectic form in terms of the variables $q$, $\dot{q}$, or
equivalently $\pi _{0}$ in the notation of Ref. \refcite{JLM}. The substitution of
equations of motion into the Lagrangian, which is normally forbidden, is
justified here in virtue of the detailed construction of Ref. \refcite{JLM},
which shows that this substitution really amounts to strongly imposing a set
of second class constraints which lead to Dirac brackets. In particular the
method proposed in Ref. \refcite{JLM} can be applied only to regular higher order
Lagrangians. The above methods are based on the projection of the Lagrangian
constraints into the phase space of momenta and coordinates, and
subsequently considering these projections as Hamiltonian constraints. But,
as has been shown in Ref. \refcite{PONS1}, Lagrangian and Hamiltonian constraints
are not directly related, and, what is more, in general they are not
projectable. In principle, this restricts the applicability of this
approach. At least, in the case of higher order gauge theories it is
necessary to implement a gauge fixing before applying these approaches.

Previous arguments made clear that HOTD theories, in particular when
considered as corrections to standard ones, require the imposition of
perturbative constraints at a given level . This implementation is closely
related to the basic question of which are the appropriate Feynman rules to
calculate a given low-energy process. To this end it is necessary to
understand the difference between HOTD theories in the context of effective
field theories and their use as HOTD theories per se. In Ref. \refcite{WEINBERG}
such a difference is explored in the context of the
Bernard-Duncan model\cite{BD}. In brief, given the full HOTD Lagrangian, the
Mathews' theorem \cite{MATTHEWS} leads to the Feynman rules as read directly from this
Lagrangian, which will imply the use of the full propagator as is proved in
Ref. \refcite{BD}. Nevertheless, as we know, this theory has all the problems
of the HOTD theories. Hence, to make it consistent as a perturbation of the
standard scalar field, we should expand the propagator in powers of the
coupling constant associated to the higher order term. With this
manipulation we obtain a perturbative expression for the propagator of the
usual scalar field, and because of the disappearance of the high-energy
poles, the ghost degrees of freedom are no more present. The important point
to be stressed here is that a perturbative expansion of the propagator of
the higher order theory provides us with a reference to test the soundness
of an effective Lagrangian.

In this paper we propose an alternative construction, which amounts
to the application of the well established Dirac method\cite{DIRAC}, to the first order
version of the HOTD theory, augmented by additional  perturbative Hamiltonian constraints.
This construction leads to the perturbative canonical formalism in a very
systematic and simple way, working from the beginning in the corresponding
phase space. The construction is purely algebraic and does not require the
solution of any system of  differential equations. Also, the method
can be directly applied to gauge theories without the need of initially
fixing the gauge, as would be the case in the constructions of Refs. \refcite{EW,JLM}.

The organization of the paper is the following. In the next section we
introduce the concept of perturbative Hamiltonian constraints, which allows
the application of the Dirac method to obtain a consistent canonical
formulation, exact to a given order in the perturbative parameter. There we also
show that the additional (perturbative) contraints can be consistently implemented in the Hamiltonian
formulation.The
method is illustrated in the third section with the construction of a
canonical formalism of arbitrary order in the perturbative parameter for the
Pais-Uhlembeck oscillator. The fourth section applies the method to the
Bernard-Duncan theory\cite{BD} with a $\varphi^{4}$ interaction, both at the
classical and quantum level. Finally, the fifth section discusses the two
particle quantum scattering in the model of the previous section in
order to compare the exact canonical theory, which contains ghosts, with the
well defined perturbative one obtained with our method. We show that our
canonical formalism recovers the results obtained by using perturbative
propagators and ruling out the unphysical states in the exact, pathological,
theory. The last section contains a summary of the main results and some
general comments. In the Appendix we consider a simple model to illustrate how our method compares with
the proposal in Ref. \refcite{EW}.

\section{Extended Dirac approach: Hamiltonian perturbative constraints}

Any HOTD Lagrangian can be rewritten as a first order one by introducing an
adequate number of auxiliary variables, via Lagrange multipliers. Once this
is done we can calculate the corresponding momenta using the usual definition. Some of these
relations allow us to write a restricted set of velocities as functions of
the coordinates and the momenta, while others yield constraints. The usual
procedure to consistently define the dynamics is the use of the Dirac
approach, and in this way we obtain the canonical formalism, which in the
case of a HOTD theory is plagued with several pathologies. In our case, the
HOTD terms turn out to be scaled by a parameter $\gamma$, which
we assume small, in such a way that the HOTD contributions are considered as
perturbations over a standard theory. Here it is possible to go directly to
a perturbative Hamiltonian construction, because some of the relations that
come from the definition of the momenta and the constraints are
inhomogeneous in the perturbative parameter. These relations, multiplied by
a power of $\gamma$, provide perturbative constraints valid up to a given
power of $\gamma$. These new constraints can be treated as Hamiltonian
ones, and added to the set of original primary constraints. Thus,
besides those primary constrains generated by the definition of the momenta,
this approach requires the introduction of additional primary
constraints further imposed according to the order of $\gamma$ to which we
decide to incorporate the HOTD corrections. From now on the
procedure follows as in the usual Dirac analysis of constrained systems.

In the following we show that this is a consistent way of considering such
constraints. To make this point clear, let us assume that we start with a
Lagrangian $L=L(q,\dot{q})$, where $p=\partial L/\partial {\dot q}$, and that
we want to introduce an external phase-space constraint, i.e. one not
generated by the definition of the momenta, $f\left( q,p\right) =0$. The
most straightforward way of doing this is by reformulating the theory in an
enlarged space $\left( q,p\right) $, where the auxiliary variable $p$
corresponds to the momentum.\ If the second order Lagrangian is regular, the
procedure for this enlargement is given in Ref. \refcite{LANCZOS}, and leads to
a well known first order Lagrangian of the form
\begin{equation}
L=p_{i}\dot{q}^{i}-H\left( q,p\right) ,
\end{equation}
where $H(q,p)$ is the corresponding Hamiltonian. Here we can impose the
additional constraint by using a Lagrange multiplier. Thus we set
\begin{equation}
L=p_{i}\dot{q}^{i}-H\left( q,p\right) +\lambda f(q,p).
\end{equation}
Next we apply the Dirac procedure to this Lagrangian, with $q^{i},p_{i}$ and
$\lambda$ considered as coordinates in the extended space. The definition of
the momenta yields three primary constraints%
\begin{equation}
\pi_{qi}-p_{i}\simeq0,\ \ \ \pi_{p}^{i}\simeq0,\ \ \ \ \pi_{\lambda}\simeq0,
\label{CONSTREXT}
\end{equation}
with the extended Hamiltonian%
\begin{equation}
H_{p}=H\left( q,p\right) -\lambda f(q,p)+u^{i}\left( \pi_{qi}-p_{i}\right)
+v_{i}\pi_{p}^{i}+w\pi_{\lambda},
\end{equation}
where $u^{i},v_{i},w$ are arbitrary functions. The consistency under time
evolution of the primary constraints fixes two of the arbitrary functions%
\begin{align}
\left\{ \pi_{qi}-p_{i},H\right\} & \simeq0\rightarrow v_{i}=-\frac {\partial
H}{\partial q_{i}}+\lambda\frac{\partial f}{\partial q_{i}}, \\
\left\{ \pi_{p}^{i},H\right\} & \simeq0\rightarrow u^{i}=\frac{\partial H}{%
\partial p_{i}}-\lambda\frac{\partial f}{\partial p_{i}},
\end{align}
where $\simeq$ indicates a weak equation. These equations generate a
secondary constraint%
\begin{equation}
\left\{ \pi_{\lambda},H\right\} \simeq0\rightarrow f\simeq0.
\end{equation}
Thus, at this level we have the Hamiltonian%
\begin{equation}
H_{D}=H\left( q,p\right) -\lambda f(q,p)+\left( \frac{\partial H}{\partial
p_{i}}-\lambda\frac{\partial f}{\partial p_{i}}\right) \left(
\pi_{qi}-p_{i}\right) -\left( \frac{\partial H}{\partial q^{i}}-\lambda
\frac{\partial f}{\partial q^{i}}\right) \pi_{p}^{i}+w\pi_{\lambda},
\end{equation}
with the set of constraints%
\begin{equation}
\pi_{qi}-p_{i}\simeq0,\;\;\;\pi_{p}^{i}\simeq0,\;\;\;\pi_{\lambda}\simeq0,\;%
\;\;\;f\simeq0.
\end{equation}

The Poisson bracket of an arbitrary function $M(q,p)$ with $H_{D}$ can be
simply written as%
\begin{equation}
\left\{ M,H_{D}\right\} =
%\frac{\partial M}{\partial q^{i}}\frac{\partial H}{%
%\partial p_{i}}-\frac{\partial M}{\partial p_{i}}\frac{\partial H}{\partial
%q^{i}}-\lambda\left( \frac{\partial M}{\partial q^{i}}\frac{\partial f}{%
%\partial p_{i}}-\frac{\partial M}{\partial p_{i}}\frac{\partial f}{\partial
%q^{i}}\right) =
\left\{ M,H\right\} ^{\prime}-\lambda\left\{ M,f\right\}
^{\prime},
\end{equation}
where $\left\{ ,\right\} ^{\prime}$\textbf{\ }is the Poisson bracket in the $%
\left( q,p\right) $ subspace. Thus, the consistency condition for the
secondary constraint $f$ is%
\begin{equation}
\left\{ f,H_{D}\right\} =\left\{ f,H\right\} ^{\prime}.
\end{equation}

If $\left\{ f,H\right\} ^{\prime}\simeq0$ there are no more constraints, and
hence $\pi_{\lambda}$ is a first class constraint. It generates an orbit of
equivalent configurations and we can choose any point of this orbit by
imposing a gauge fixing. It is convenient to take $\lambda=0$. Besides this,
the first two constraints in Eq. (\ref{CONSTREXT}) are second class. We
finally get, in the reduced space defined by the gauge fixing and these two
second class constraint, that the dynamics is described by the usual
Hamiltonian $H_{D}^{R}=H\left( q,p\right)$,
with only the constraint we want to impose, $f(q,p)\simeq0$.

If $\left\{ f,H\right\} ^{\prime}$ is not weakly zero there is a new
constraint%
\begin{equation}
\left\{ \left\{ f,H_{D}\right\} ,H_{D}\right\} =\left\{ \left\{ f,H\right\}
^{\prime},H_{D}\right\} =\left\{ \left\{ f,H\right\} ^{\prime},H\right\}
^{\prime}-\lambda\left\{ \left\{ f,H\right\} ^{\prime },\;f\right\}
^{\prime},
\end{equation}
whose consistency condition fixes the remaining arbitrary function $w$%
\begin{align}
&\left\{ \left\{ \left\{ f,H\right\} ^{\prime},H\right\}
^{\prime}-\lambda\left\{ \left\{ f,H\right\} ^{\prime},f\right\}
^{\prime},H_{D}\right\}=\notag\\
&\left\{ \left\{ \left\{ f,H\right\} ^{\prime
},H\right\} ^{\prime},H\right\} ^{\prime}  
-\lambda\left\{ \left\{ \left\{ f,H\right\} ^{\prime},H\right\}
^{\prime},f\right\} ^{\prime}-\lambda\left\{ \left\{ \left\{ f,H\right\}
^{\prime},f\right\} ^{\prime},H\right\}  \notag \\
& +\lambda^{2}\left\{ \left\{ \left\{ f,H\right\} ^{\prime},f\right\}
^{\prime},f\right\} ^{\prime}-w\left\{ \left\{ f,H\right\} ^{\prime
},f\right\} ^{\prime}.
\end{align}

Now we have the following set of second class constraints%
\begin{align}
\pi_{qi}-p_{i}\simeq0,\;\ \pi_{p}^{i}\simeq0,\;\ \pi_{\lambda}\simeq0, & \;\
f\simeq0,  \label{CR1} \\
\left\{ \left\{ f,H\right\} ^{\prime},H\right\} ^{\prime}-\lambda\left\{
\left\{ f,H\right\} ^{\prime},f\right\} ^{\prime} & \simeq0.  \label{CR2}
\end{align}
Using the first three constraints in (\ref{CR1})\ and the last one in (\ref%
{CR2}) to partially reduce the phase space we get%
\begin{equation}
H_{D}^{R}=H\left( q,p\right) -\frac{\left\{ \left\{ f,H\right\} ^{\prime
},H\right\} ^{\prime}}{\left\{ \left\{ f,H\right\} ^{\prime},f\right\} }f,
\label{HAM2CASE}
\end{equation}
with $f\simeq0$.
In both cases, $\left\{ f,H\right\} ^{\prime}=0$ or $\left\{ f,H\right\}
^{\prime}\neq0$, the final result corresponds to considering the original
Hamiltonian in the $\left( q,p\right) $ phase space, plus the external
constraint as a Hamiltonian one in the Dirac approach.%
\begin{equation}
H\left( q,p\right) \ \ \ \text{plus}\ \ \ \ f\left( q,p\right) =0\quad
\leftrightarrow \quad H_{Dirac}=H\left( q,p\right) +u\;f\left( q,p\right) .
\end{equation}

This construction shows that in the regular case the inclusion of external
constraints involving momenta is equivalent to considering these constraints
as primary ones in the Dirac formalism. The key for this demonstration is
the construction of a first order Lagrangian with a $\left( q,p\right) $
configuration space. In the case of a singular Lagrangian, this first order
Lagrangian with a $\left( q,p\right) $ configuration space can also be
constructed following an extension of the scheme given in Lanczos\cite%
{LANCZOS}, developed in Ref. \refcite{MM}. Using this construction and
following the preceding discussion, we can show that in any case external
constraints involving coordinates and momenta can be incorporated as primary
Hamiltonian constraints in the framework of the Dirac method. In particular,
this justifies the insertion of the perturbative constraints in the Dirac
formalism, together with the ones naturally generated by the definition of
the momenta.

In the following we illustrate the proposal by applying it to two systems:
(i) the Pais-Uhlenbeck oscillator and (ii) the higher order scalar theory
discussed in Ref. \refcite{BD}, plus a self-interacting term.

\section{The Pais-Uhlenbeck oscillator}

We implement the proposal in the framework of the Pais-Uhlenbeck oscillator,
which is a regular system. For arbitrary theories described by a first-order
Lagrangian we have a well-defined consistent canonical approach, given by
the Dirac method. For this reason, our first step will be to rewrite any
higher order theory in terms of a first order Lagrangian by introducing
auxiliary degrees of freedom together with the corresponding Lagrange
multipliers. In this way we obtain a first order singular Lagrangian to
which we can apply the Dirac method to obtain the canonical formalism. When
we eliminate the auxiliary variables in this canonical formalism, by
implementing the corresponding second class Hamiltonian constraints, we
recover the standard Ostrogradsky approach \cite{JP,NESTERENKO}. An
alternative approach in the case of the Pais-Uhlenbeck oscillator based on
complex canonical transformations plus subsequent reality conditions is
presented in Ref. \refcite{VERGARA}.

The second-order Lagrangian defining the Pais-Uhlenbeck oscillator is%
\begin{equation}
L=\frac{\dot{x}^{2}}{2}-\frac{\omega^{2}x^{2}}{2}-\frac{\gamma}{2}\ddot{x}%
^{2}.  \label{LAGPU}
\end{equation}
The introduction of the additional degree of freedom $z$, via the
corresponding constraint, leads to the first-order Lagrangian%
\begin{equation}
L=\frac{z^{2}}{2}-\frac{\omega^{2}x^{2}}{2}-\frac{1}{2}\gamma\dot{z}%
^{2}+\lambda\left( z-\dot{x}\right) ,
\end{equation}
with coordinates $x,z,\lambda$. The corresponding equations of motion are%
\begin{align}
z+\gamma\ddot{z}+\lambda & =0, \\
-\omega^{2}x+\dot{\lambda} & =0, \\
z-\dot{x} & =0.
\end{align}
They imply only one perturbative primary constraint at order $\gamma^{n}$%
\begin{equation}
\gamma^{n}\left( z+\lambda\right) =0,
\end{equation}
which leads to the secondary constraint%
\begin{equation}
\gamma^{n}\left( \dot{z}+\dot{\lambda}\right) =\gamma^{n}\left( \dot{z}%
+\omega^{2}x\right) =0.
\end{equation}

One can easily verify that eliminating $z$, and $\lambda$ from the resulting
equations of motion produces the equation of motion for $x$ directly
obtained from Eq. (\ref{LAGPU}). Now that we have a first-order theory, we
can construct the canonical formalism using the Dirac approach. The
corresponding momenta $p_{k}:p,\pi,\kappa$ are%
\begin{align}
p & =\frac{\partial L}{\partial\dot{x}}=-\lambda\rightarrow\;\sigma
_{2}=\lambda+p\simeq0,  \label{momenta1} \\
\pi & =\frac{\partial L}{\partial\dot{z}}=-\gamma\dot{z},  \label{momenta2}
\\
\kappa & =\frac{\partial L}{\partial\dot{\lambda}}=0 \, \rightarrow \, \sigma
_{1}=\kappa\simeq0.  \label{momenta3}
\end{align}
This set of equations gives the transformation of the momenta into
coordinates and velocities. The equation (\ref{momenta2}) defines a
perturbative constraint of order $n$%
\begin{equation}
\Phi^{n}=\gamma^{n}\pi\simeq0,
\end{equation}
which means that we take $\gamma^{m}=0,\;\;m>n$. In this way the velocity $%
\dot{z}$ is expressed in terms of the associated momentum. There are two
primary constraints, and one perturbative primary constraint is imposed. The
primary Hamiltonian is%
\begin{equation}
H=-\frac{\pi^{2}}{2\gamma}-\frac{z^{2}}{2}+\frac{\omega^{2}x^{2}}{2}-\lambda
z+u\left( p+\lambda\right) +v\gamma^{n}\pi+w\kappa.
\end{equation}
The time evolution of the exact primary constraints and the perturbative one
yields%
\begin{align}
\left\{ p+\lambda,H\right\} & =-\omega^{2}x+w \ \rightarrow \ w=\omega^{2}x, \\
\left\{ \kappa,H\right\} & =z-u \ \rightarrow \ u=z, \\
\gamma^{n}\left\{ \pi,H\right\} & =\gamma^{n}\left( z+\lambda\right)
\rightarrow\Psi^{n}=\gamma^{n}\left( z+\lambda\right) \simeq\gamma
^{n}\left( z-p\right) \simeq0.
\end{align}
Two arbitrary functions are fixed, and a secondary constraint is generated.
The Hamiltonian becomes%
\begin{equation}
H=-\frac{\pi^{2}}{2\gamma}-\frac{z^{2}}{2}+\frac{\omega^{2}x^{2}}{2}%
+z_{1}p+v\gamma^{n}\pi+\omega^{2}x\kappa.  \label{HAM3}
\end{equation}
The consistency conditions for the secondary constraints generates a pair of
towers of constraints of decreasing order in $\gamma$, starting from the
initial $\Phi^{n},\Psi^{n}$%
\begin{equation}
\frac{d\Phi^{n}}{dt}\simeq0\;\rightarrow\Psi^{n},\;\;\frac{d\Psi^{n}}{dt}%
\simeq0\rightarrow\;\Phi^{n-1},\;\;\frac{d\Phi^{n-1}}{dt}\simeq
0\rightarrow\;\Psi^{n-1},\;....  \label{TIMEDER}
\end{equation}
according to%
\begin{equation}
\left\{ H,\Phi^{n-m}\right\} =\Psi^{n-m},\;\;\;\ \left\{ H,\Psi
^{n-m}\right\} =\Phi^{n-m-1}.
\end{equation}

The general expression for the constraints can be written in terms of the
Fibonacci polynomials \cite{FIBO}%
\begin{equation}
F_{m}\left( y\right) =\sum_{i=0}^{\left[ \left( m+1\right) /2\right] }\left(
-1\right) ^{i}\left(
\begin{array}{c}
m+1-i \\
i%
\end{array}
\right) y^{i},\;\;F_{0}\left( y\right) =1,\;\;\;\;F_{-1}(y)=1,
\end{equation}
where $\left[ a\right] $ means the integer part of $a$, and are%
\begin{align}
\Phi^{n} & =\gamma^{n}\pi, \\
\Psi^{n-m} & =\gamma^{n-m}\left( zF_{m}\left( \gamma\omega^{2}\right)
-pF_{m-1}\left( \gamma\omega^{2}\right) \right) \qquad \ \ \ \ \ \ \ \ \ \ \
\ \ 0\leq m\leq n, \\
\Phi^{n-m} & =\gamma^{n-m}\left( \pi F_{m-1}\left( \gamma\omega^{2}\right)
-x\ \gamma\omega^{2}F_{m-2}\left( \gamma\omega^{2}\right) \right)
\qquad1\leq m\leq n.
\end{align}
The Fibonacci polynomials satisfy%
\begin{equation}
F_{m+1}(y)=F_{m}(y)-yF_{m-1}(y),\;\;\;\;F_{0}(y)=1,\;\;\;F_{-1}(y)=1,
\end{equation}
which imply the recurrence relations%
\begin{align}
\Psi^{n-m} & =\gamma\Psi^{n-m-1}+\omega^{2}\Psi^{n-m+1}\qquad0\leq m\leq n,\
\ \ \ \Psi^{-1}=\Psi^{n+1}=0, \\
\Phi^{n-m} & =\gamma\Phi^{n-m-1}+\omega^{2}\Phi^{n-m+1}\qquad1\leq m\leq n,\
\ \ \ \Phi^{-1}=\Phi^{n+1}=0,
\end{align}
for the constraints.

Using these recurrence relations we can express the complete set of
constraints as proportional to the last ones, $\Psi^{0}$ and $\Phi^{0}$.
Thus we finally get%
\begin{align}
\Psi^{m} & =\omega^{-2m}F_{m-1}\left( \gamma\omega^{2}\right) \Psi ^{0},
\label{REC1} \\
\Phi^{m} & =\omega^{-2m}F_{m-1}\left( \gamma\omega^{2}\right) \Phi ^{0},
\label{REC2}
\end{align}
with%
\begin{align}
\Psi^{0} & =zF_{n}\left( \gamma\omega^{2}\right) -pF_{n-1}\left(
\gamma\omega^{2}\right) , \\
\Phi^{0} & =\pi F_{n-1}\left( \gamma\omega^{2}\right) -x\ \gamma\omega
^{2}F_{n-2}\left( \gamma\omega^{2}\right) .
\end{align}

It is important to emphasize that the final perturbative constraints $\Psi
^{0},\;\Phi^{0}$ imply all the remaining ones up to $\Psi^{n},\;\Phi^{n}$,
and for this they are taken as the independent perturbative constraints to
the order considered.

For simplicity, the illustration of the complete Hamiltonian construction is
restricted here to the case $n=1$, defined by the external perturbative
constraint $\gamma\pi\simeq0$. The non-perturbative constraints (\ref%
{momenta1},\ref{momenta3}) remain valid for all orders. In this case the two
independent perturbative constraints become%
\begin{align}
\Phi^{0} & \equiv\chi_{1}=\pi-\gamma\omega^{2}x, \\
\Psi^{0} & \equiv\chi_{2}=z-\left( 1+\gamma\omega^{2}\right) p.
\end{align}
The system is now second class, and it is very easy to compute from Eq. (\ref%
{HAM3}) the Dirac Hamiltonian in the constrained space%
\begin{equation}
H=\frac{1}{2}p^{2}+\frac{1}{2}\omega^{2}\left( 1-\gamma\omega^{2}\right)
x^{2},
\end{equation}
together with the corresponding Dirac bracket%
\begin{align}
\left\{ p,x\right\} _{D} & =1-\left( 1+\gamma\omega^{2}\right) \left(
1+\gamma\omega^{2}\right) \gamma\omega^{2}\simeq\left( 1-\gamma\omega
^{2}\right) .
\end{align}
It is clear that $x$ and $p$ are not canonical conjugate variables. Instead
we can use%
\begin{equation}
x\rightarrow x,\;\;\;\;\;\;p\rightarrow\tilde{p}=\left( 1-\gamma\omega
^{2}\right) p,
\end{equation}
which satisfy%
\begin{equation}
\left\{ \tilde{p},x\right\} _{D}=\left( 1-\gamma\omega^{2}\right) \left\{
p,x\right\} _{D}\simeq1.
\end{equation}
Thus we get%
\begin{equation}
\tilde{H}=\frac{1}{2}\tilde{p}^{2}+\frac{1}{2}\omega^{2}\left( 1+\omega
^{2}\gamma\right) x^{2},
\end{equation}
to first order in $\gamma$. Now we can read the frequency of the oscillator
in the Hamiltonian, which coincides with that obtained from the perturbative
equation of motion to the order considered.

\section{Higher order scalar field theory: Bernard-Duncan field with a $%
\protect\varphi^{4}$ interaction}

We now consider a higher order field theory, given by a Lagrangian density
which is a self-interacting generalization of the Pais-Uhlenbeck oscillator%
\begin{equation}
\mathcal{L}=-\frac{1}{2}\varphi\square\varphi-\frac{1}{2}\varphi
m^{2}\varphi+\frac{\gamma}{2}\varphi\square^{2}\varphi-\frac{\alpha}{4}%
\varphi ^{4},\;\;\;\;\square=\partial^{\mu}\partial_{\mu}.  \label{BL}
\end{equation}
The HOTD equation of motion is%
\begin{equation}
\square\varphi+m^{2}\varphi-\gamma\square^{2}\varphi+\alpha\varphi ^{3}=0.
\label{EMB}
\end{equation}
Using an iterative procedure
for the Pais-Uhlenbeck model,\ the above equation reduces to the following
second order equation%
\begin{equation}
\square\varphi+m^{2}\left( 1-\gamma m^{2}\right) \varphi+\left( 1-4\gamma
m^{2}\right) \alpha\varphi^{3}-3\gamma\alpha^{2}\varphi^{5}+6\gamma
\alpha\varphi\left( \partial_{\mu}\varphi\partial^{\mu}\varphi\right) =0,
\label{FOEQMO}
\end{equation}
to first order in $\gamma$.

We deal with this model following the prescription in the previous
subsection, so that the Lagrangian (\ref{BL}) is first rewritten as a
first-order one\ with respect to the time derivatives, by introducing an
auxiliary variable $\psi=\partial_{0}\varphi$ with a Lagrange multiplier $%
\lambda$%
\begin{equation}
\mathcal{L}^{\prime}=\frac{1}{2}\psi^{2}-\frac{1}{2}\varphi\left(
m^{2}-\Delta\right) \varphi+\gamma\psi\Delta\psi+\frac{\gamma}{2}%
\varphi\Delta^{2}\varphi-\frac{\alpha}{4}\varphi^{4}+\lambda\left( \dot{%
\varphi}-\psi\right) +\frac{\gamma}{2}\dot{\psi}^{2}.
\end{equation}
The definition of the canonical momenta establishes%
\begin{equation}
\pi_{\psi}=\gamma\dot{\psi},  \label{pf}
\end{equation}
and gives two Hamiltonian primary constraints%
\begin{align}
\chi_{1} & \equiv\pi_{\lambda}\simeq0,  \label{BC1} \\
\chi_{2} & \equiv\pi_{\varphi}-\lambda\simeq0,  \label{BC2}
\end{align}
In the first place we will consider this theory without any approximation in
$\gamma$, and in the next subsection it will be reformulated using the
Hamiltonian constraint approach.

\subsection{ The $\protect\gamma$-non-perturbative formalism}

Here we construct the canonical formalism using the Dirac approach. The
primary Hamiltonian density becomes%
\begin{equation}
\mathcal{H}=-\frac{1}{2\gamma}\pi_{\psi}^{2}-\frac{1}{2}\psi\left(
1-2\gamma\Delta\right) \psi+\frac{1}{2}\varphi\left( m^{2}-\Delta
-\gamma\Delta^{2}\right) \varphi+\frac{\alpha}{4}\varphi^{4}+\lambda
\psi+w\left( \pi_{\varphi}-\lambda\right) +u\pi_{\lambda},
\end{equation}
where $w$ and $u$ are arbitrary functions. The consistency conditions for
the primary constraints only fix two of the arbitrary functions%
\begin{align}
\left\{ \pi_{\varphi}-\lambda,\;H\right\} & =-\left( m^{2}-\Delta
-\gamma\Delta^{2}\right) \varphi-\alpha\varphi^{3}-u\simeq 0
, \\
\left\{ \pi_{\lambda},H\right\} & =-\psi +w\simeq 0,
\end{align}
where $H=\int d^{3}x\mathcal{H}$. No further constraints are generated, and
thus the Hamiltonian density becomes%
\begin{align}
\mathcal{H}=\frac{1}{2\gamma}\pi_{\psi}^{2}-\psi\left(
\frac{1}{2}+\gamma\Delta\right) \psi+\left(\frac{\varphi}{2}-\pi_{\lambda}\right)\left( m^{2}-\Delta
-\gamma\Delta^{2}\right) \varphi+\frac{\alpha}{4}\varphi^{4}+\psi\pi
_{\varphi}-\alpha\pi_{\lambda}\varphi^{3}.
\end{align}
The primary constraints are second class%
\begin{equation}
\left\{ \chi_{1}\left( x\right) ,\chi_{2}\left( x^{\prime}\right) \right\}
=\delta\left( x-x^{\prime}\right) ,
\end{equation}
and therefore we can directly use the reduced Hamiltonian density%
\begin{equation}
\mathcal{\tilde{H}}=\frac{1}{2\gamma}\pi_{\psi}^{2}+\psi\pi_{\varphi}-\frac {%
1}{2}\psi\left( 1+2\gamma\Delta\right) \psi+\frac{1}{2}\varphi\left(
m^{2}-\Delta-\gamma\Delta^{2}\right) \varphi+\frac{\alpha}{4}\varphi ^{4},
\label{HAMTILDE}
\end{equation}
with the Dirac brackets%
\begin{align}
\left\{ \varphi\left( x,t\right) ,\pi_{\varphi}\left( y,t\right) \right\}
_{D} & =\left\{ \varphi\left( x,t\right) ,\pi_{\varphi}\left( y,t\right)
\right\} =\delta\left( x-y\right) , \\
\left\{ \psi\left( x,t\right) ,\pi_{\psi}\left( y,t\right) \right\} _{D} &
=\left\{ \psi\left( x,t\right) ,\pi_{\psi}\left( y,t\right) \right\}
=\delta\left( x-y\right) .
\end{align}
The theory discussed by Bernard and Duncan corresponds to $\alpha=0$. We
will consider now this case, which can be treated in an exact way. To do
this, it is useful to decompose the field $\varphi$ in modes with well
defined frequency and with covariant normalization%
\begin{align}
\varphi\left( x,t\right) =\int d^{3}p&\left( \frac{1}{\sqrt{2\omega_{p}}}%
\left( a_{p}e^{i\left( \mathbf{p\cdot x}-\omega_{p}t\right)
}+a_{p}^{\dagger}e^{-i\left( \mathbf{p\cdot x}-\omega_{p}t\right) }\right)\right .\notag\\
&\left .+
\frac{1}{\sqrt{2\Omega_{p}}}\left( b_{p}e^{i\left( \mathbf{p\cdot x}%
-\Omega_{p}t\right) }+b_{p}^{\dagger}e^{-i\left( \mathbf{p\cdot x}%
-\Omega_{p}t\right) }\right) \right) .  \label{PHIDESCOMP}
\end{align}
The characteristic frequencies result%
\begin{align}
\omega_{p} & =\left(\mathbf{p}^{2}+\frac{ \sqrt{1+4\gamma m^{2}}%
-1 }{2\gamma}\right)^{1/2}, \\
\Omega_{p} & =\left(\mathbf{p}^{2}-\frac{ \sqrt{1+4\gamma m^{2}}%
+1 }{2\gamma}\right)^{1/2}.
\end{align}
When $-1 < 4m^{2}\gamma<0$, both frequencies are real. When\ $%
4m^{2}\gamma<-1 $ both frequencies have an imaginary part. In
the case $\gamma>0$, both terms $\left( \sqrt{1+4\gamma m^{2}}\pm1\right)$,
are positive so that $\omega_{p}$ is always real, while $\Omega_{p}$ turns
out to be imaginary when $\mathbf{p}^{2}<\left( \sqrt{1+4\gamma m^{2}}%
+1\right) /2\gamma$. Those sectors of the theory with complex frequencies
give rise to runaway solutions.

From the canonical brackets%
\begin{equation}
\left\{ \varphi\left( x,t\right) ,\pi_{\varphi}\left( x^{\prime},t\right)
\right\} =i\delta\left( x-x^{\prime}\right) ,\;\;\;\;\;\;\left\{ \psi\left(
x,t\right) ,\pi_{\psi}\left( x^{\prime},t\right) \right\} =i\delta\left(
x-x^{\prime}\right) ,
\end{equation}
together with the following equations of motion arising from (\ref{HAMTILDE})%
\begin{equation}
\dot{\varphi}=\psi,\;\;\dot{\psi}=\frac{1}{\gamma}\pi_{\psi},\;\pi_{\varphi
}=\left( 1+2\gamma\Delta\right) \psi-\dot{\pi}_{\psi}\;,
\end{equation}
which allow us to express the remaining fields $\psi,\pi_{\psi}$\ and $%
\pi_{\varphi}$ in terms of $a_{p},a_{p^{\prime}}^{\dagger},b_{p},b_{p^{%
\prime}}^{\dagger}$, we get%
\begin{align}
\left\{ a_{p},a_{p^{\prime}}^{\dagger}\right\} & =\frac{1}{\sqrt{1+4\gamma
m^{2}}}\delta^{3}\left( \mathbf{p}-\mathbf{p}^{\prime}\right) , \\
\left\{ b_{p},b_{p^{\prime}}^{\dagger}\right\} & =-\frac{1}{\sqrt {1+4\gamma
m^{2}}}\delta^{3}\left( \mathbf{p}-\mathbf{p}^{\prime}\right) , \\
\left\{ a_{p},a_{p^{\prime}}\right\} & =\left\{ a_{p}^{\dagger
},a_{p^{\prime}}^{\dagger}\right\} =\left\{ b_{p},b_{p^{\prime}}\right\}
=\left\{ b_{p}^{\dagger},b_{p^{\prime}}^{\dagger}\right\} =0,
\end{align}
which is the same result obtained in Ref. \refcite{BD}, with $\gamma
\rightarrow-\gamma$ and a different definition for the auxiliary field. Let
us recall that any Poisson bracket between $a_{p},a_{p^{\prime}}^{\dagger}$\
and $b_{p},b_{p^{\prime}}^{\dagger}$ is zero. With the above normalization
the number operator is%
\begin{equation}
N=\sqrt{1+4\gamma m^{2}}\int d^{3}p\;\left[ a_{p}^{\dagger}a_{p}+b_{p}^{%
\dagger}b_{p}\right] .
\end{equation}

In terms of the momentum space fields the Hamiltonian becomes%
\begin{equation}
H=\frac{1}{2}\int d^{3}p\left( \left( a_{p}a_{p}^{\dagger}+a_{p}^{\dagger
}a_{p}\right) \omega_{p}+\left(
b_{p}b_{p}^{\dagger}+b_{p}^{\dagger}b_{p}\right) \Omega_{p}\right) .
\end{equation}
From here the canonical quantization is straightforward. The algebra of the
creation and annihilation operators is%
\begin{align}
\left[ a_{p},a_{p^{\prime}}^{\dagger}\right] & =\frac{1}{\sqrt{1+4\gamma
m^{2}}}\delta^{3}\left( \mathbf{p}-\mathbf{p}^{\prime}\right) \\
\left[ b_{p},b_{p^{\prime}}^{\dagger}\right] & =-\frac{1}{\sqrt{1+4\gamma
m^{2}}}\delta^{3}\left( \mathbf{p}-\mathbf{p}^{\prime}\right)  \label{WRONGS}
\\
\left[ a_{p},a_{p^{\prime}}\right] & =\left[ a_{p}^{\dagger},a_{p^{%
\prime}}^{\dagger}\right] =\left[ b_{p},b_{p^{\prime}}\right] =\left[
b_{p}^{\dagger},b_{p^{\prime}}^{\dagger}\right] =0
\end{align}
and the normal ordered Hamiltonian operator results%
\begin{equation}
\hat{H}_{N}=\int d^{3}p\left( a_{p}^{\dagger}a_{p}\omega_{p}+b_{p}^{\dagger
}b_{p}\Omega_{p}\right) ,
\end{equation}
which acts on one-particle states according to%
\begin{align}
\hat{H}_{N}a_{k}^{\dagger}\left\vert 0,0\right\rangle & =\frac{\omega_{k}}{%
\sqrt{1+4\gamma m^{2}}}a_{k}^{\dagger}\left\vert 0,0\right\rangle
\label{1PA} \\
\hat{H}_{N}b_{k}^{\dagger}\left\vert 0,0\right\rangle & =-\frac{\Omega_{k}}{%
\sqrt{1+4\gamma m^{2}}}b_{k}^{\dagger}\left\vert 0,0\right\rangle
\label{1PB}
\end{align}
Due to the minus sign in Eqs. (\ref{WRONGS},\ref{1PB}), the requirement of
energy positivity implies a negative metric for states containing type $b$
particles, spoiling physical unitarity and a consistent probabilistic
interpretation. A normalized state with $n$ particles ($s$ particles of type
$a$, with momenta $p_{1},...,p_{s}$ and $\left( n-s\right)$ particles of
type $b$\ with momenta $p_{s+1},...,p_{n}$) in the $\alpha=0$\ case is given
by%
\begin{equation}
\left\vert \Phi_{a,b}\left( p_{1,}\text{ }p_{2},...,p_{n}\right)
\right\rangle =\left( 1+4\gamma m^{2}\right) ^{\frac{n}{4}}\left( \Pi
_{i=1}^{s}a_{p_{i}}^{\dagger}\right) \left( \Pi_{i=s+1}^{n}b_{p_{j}}^{\dag
}\right) \left\vert 0,0\right\rangle .  \label{ns}
\end{equation}

The expression (\ref{1PB}) shows that the minus sign in the commutation
relations (\ref{WRONGS}) leads to negative contributions for the energy from
the excitations of type $b$.

\subsection{The perturbative Hamiltonian constraint approach}

Here we go back to the self-interacting case $\alpha\neq0$. The expression
for $\pi_{\psi}$, equation (\ref{pf}), is non homogenous in $\gamma$, and
hence generates a Hamiltonian perturbative constraint of order $\gamma^{n}$%
\begin{equation}
\gamma^{n}\pi_{\psi}=0,
\end{equation}
where $n$ is the order of the perturbative approximation we want to achieve.
Thus, the primary Hamiltonian density for a perturbative approach of order $%
n $ in $\gamma$ is%
\begin{align}
\mathcal{H} & =\frac{1}{2\gamma}\pi_{\psi}^{2}-\frac{1}{2}\psi\left(
1+2\gamma\Delta\right) \psi+\frac{1}{2}\varphi\left( m^{2}-\Delta
-\gamma\Delta^{2}+\frac{\alpha}{2}\varphi^{2}\right) \varphi  \notag \\
& +\psi\pi_{\varphi}-\pi_{\lambda}\left( m^{2}-\Delta-\gamma\Delta
^{2}+\alpha\varphi^{2}\right) \varphi+v\gamma^{n}\pi_{\psi}
\end{align}
and the full set of constraints,\ primary and secondary, are given by (\ref%
{BC1}) and (\ref{BC2}) together with the chain%
\begin{align}
\Phi_{0} & =\gamma^{n}\pi_{\psi}\simeq0 \\
\Psi_{0} & =\gamma^{n}\left( \psi-\pi_{\varphi}\right) \simeq0 \\
\Phi_{1} & =\gamma^{n-1}\left( \gamma\left( m^{2}-\Delta+\alpha\varphi
^{2}\right) \varphi+\pi_{\psi}\right) \simeq0 \\
\Psi_{1} & =\gamma^{n-1}\left[ \left( \gamma\left(
\Delta+m^{2}+3\alpha\varphi^{2}\right) +1\right) \psi-\pi_{\varphi}\right]
\simeq0 \\
&
\begin{array}{c}
\cdot \\
\cdot \\
\cdot%
\end{array}
\notag
\end{align}
which are obtained by requiring
\begin{equation}
\left\{ H,\Phi^{n-m}\right\} =\Psi^{n-m},\;\;\;\ \left\{ H,\Psi
^{n-m}\right\} =\Phi^{n-m-1}.
\end{equation}
As a simple illustration, let us consider the case $n=1$. The above chain
produces the additional perturbative constraints%
\begin{align}
\Phi_{1} & =\gamma\left( m^{2}-\Delta+\alpha\varphi^{2}\right) \varphi
+\pi_{\psi}=0  \label{PERT1} \\
\Psi_{1} & =\left( \gamma\left( \Delta+m^{2}+3\alpha\varphi^{2}\right)
+1\right) \psi-\pi_{\varphi}=0  \label{PERT2}
\end{align}
which have to be considered together with the original ones (\ref{BC1})\ and
(\ref{BC2}).

The set \ (\ref{BC1}),\ (\ref{BC2}), (\ref{PERT1})\ and (\ref{PERT2})\
corresponds to four second class constraints, leading to the Dirac brackets%
\begin{align}
\left\{ \varphi,\pi_{\varphi}\right\} _{D}=\left\{ \varphi,\pi_{\varphi
}\right\} -\left\{ \varphi,\Psi_{1}\right\} C_{\Psi_{1}\Phi_{1}}^{-1}\left\{
\Phi_{1},\pi_{\varphi}\right\} &-\left\{ \varphi,\chi _{2}\right\}
C_{\chi_{2}\Phi_{1}}^{-1}\left\{ \Phi_{1},\pi_{\varphi }\right\}\notag\\
 &-\left\{
\varphi,\chi_{2}\right\} C_{\chi_{2}\Psi_{1}}^{-1}\left\{
\Psi_{1},\pi_{\varphi}\right\} ,
\end{align}
where $C_{ab}$ is the standard matrix of the Poisson brackets among the
second class constraints. The final result in the $\left( \varphi
,\pi_{\varphi}\right) $ reduced phase space is%
\begin{equation}
\left\{ \varphi,\pi_{\varphi}\right\} _{D}=1-\gamma\left(
m^{2}-\Delta+3\alpha\varphi^{2}\right) ,  \label{DB}
\end{equation}
together with the Hamiltonian density%
\begin{align}
\mathcal{H} & =\frac{1}{2}\pi_{\varphi}\left( 1-2\gamma\Delta\right)
\pi_{\varphi}+\frac{1}{2}m^{2}\left( 1+\gamma m^{2}\right) \varphi^{2}-\frac{%
1}{2}\left( 1+2\gamma m^{2}\right) \varphi\Delta\varphi%
\left( 1+4\gamma m^{2}\right) \varphi^{4}  \notag \\
& +\frac{\alpha }{4}+\frac{1}{2}\gamma\alpha^{2}\varphi^{6}-\gamma\alpha\left( 3\varphi
^{2}\left( \triangledown\varphi\right) ^{2}+2\varphi^{3}\Delta
\varphi\right) .  \label{HAMB1}
\end{align}
The final Dirac bracket (\ref{DB}) is not canonical. To express the
Hamiltonian density in terms of canonical variables, maintaining the
original field $\varphi$, it is necessary to apply a non canonical
transformation%
\begin{align}
\tilde{\varphi} & =\varphi, \\
\tilde{\pi}_{\varphi} & =\left( 1+\gamma\left( m^{2}-\Delta+3\alpha
\varphi^{2}\right) \right) \pi_{\varphi},
\end{align}
such that%
\begin{equation}
\left\{ \tilde{\varphi},\tilde{\pi}_{\varphi}\right\} _{D}=1+\mathcal{O}%
\left( \gamma^{2}\right) .
\end{equation}
In terms of this new momentum, which we call again $\pi_{\varphi}$ in an
abuse of notation, the Hamiltonian density (\ref{HAMB1})\ becomes%
\begin{align}
\mathcal{H} & =\frac{1}{2}\pi_{\varphi}\left( 1-2\gamma\left(
m^{2}+3\alpha\varphi^{2}\right) \right) \pi_{\varphi}+\frac{1}{2}m^{2}\left(
1+\gamma m^{2}\right) \varphi^{2}  \notag \\
& -\frac{1}{2}\left( 1+2\gamma\left( m^{2}+\alpha\varphi^{2}\right) \right)
\varphi\Delta\varphi+\frac{\alpha}{4}\left( 1+4\gamma m^{2}\right)
\varphi^{4}+\frac{1}{2}\gamma\alpha^{2}\varphi^{6}.  \label{HAMB2}
\end{align}
It is straightforward to verify that this Hamiltonian density yields the
same equation of motion (\ref{FOEQMO}) as the original Lagrangian density (%
\ref{BL}), to first order in $\gamma$. To close the discussion, we obtain
from Eq. (\ref{HAMB2})\ the effective Lagrangian density to first order in $%
\gamma$ in the configuration space $\varphi$%
\begin{align}
\mathcal{L}_{D}=\pi_{\varphi}\dot{\varphi}-\mathcal{H}=&-\left( \frac{1}{2}+\gamma
m^{2}\right) \left( \varphi\square\varphi+m^{2}\left(
1-\gamma m^{2}\right) \varphi^{2}\right . \notag\\
&\left . +\frac{\alpha}{2}\left( 1+2\gamma
m^{2}\right) \varphi^{4}+2\gamma\alpha\varphi^{3}\square \varphi+%
\alpha^{2}\gamma\varphi^{6}\right) .  \label{LAGBGAMA1}
\end{align}
This Lagrangian density has to be compared with the exact one, given by Eq. (%
\ref{BL}). The effect of the HOTD term in this effective Lagrangian has been
to produce modifications in the mass of the field $\varphi$, in the coupling
constant of the self-interaction $\varphi^{4}$, and has also generated two
new interaction terms, one of order $\gamma\varphi^{4}$ with a derivative
coupling and another of order $\gamma\varphi^{6}$. Normalizing the kinetic
term via the substitution%
\begin{equation}
\varphi\rightarrow\tilde{\varphi}=\left( 1+\gamma m^{2}\right) \varphi,
\label{REDEFPHI}
\end{equation}
we get%
\begin{equation}
\mathcal{L}_{D}=-\frac{1}{2}\tilde{\varphi}\square\tilde{\varphi}-\frac{1}{2}%
m^{2}\left( 1-\gamma m^{2}\right) \tilde{\varphi}^{2}-\frac{\alpha}{4}\tilde{%
\varphi}^{4}-\gamma\alpha\tilde{\varphi}^{3}\square\tilde{\varphi }-\frac{1}{%
2}\alpha^{2}\gamma\tilde{\varphi}^{6}.  \label{FINLAGB}
\end{equation}

Before closing this subsection it is interesting to compare the final
Lagrangian density (\ref{FINLAGB}), obtained via the perturbative
Hamiltonian constraint method, with different alternatives previously
proposed to reduce (\ref{BL}) to an effective first order form.

One such alternative is the double zero method\cite{BARKEROCONELL}. In this
case we can use de equation for motion (\ref{EMB}) to generate such a term,
which leads to%
\begin{equation}
\mathcal{\tilde{L}}_{DZ}=-\frac{1}{2}\varphi\square\varphi-\frac{1}{2}%
\varphi m^{2}\varphi+\frac{\gamma}{2}\varphi\square^{2}\varphi-\frac{\alpha}{%
4}\varphi^{4}-\frac{1}{2}\gamma\left( \square\varphi+m^{2}\varphi+\alpha
\varphi^{3}\right) ^{2},
\end{equation}
and thus the effective Lagrangian to first order in $\gamma$ results
identical to (\ref{LAGBGAMA1}), so that after the redefinition (\ref%
{REDEFPHI}) reproduces (\ref{FINLAGB}).

Another possibility is to implement an appropriate derivative transformation
in (\ref{BL}) , to first order in $\gamma$, given by%
\begin{equation}
\varphi\rightarrow\varphi=\left( \tilde{\varphi}+\frac{1}{2}\gamma \square%
\tilde{\varphi}\right) ,
\end{equation}
which yields the following first-order Lagrangian density to order $\gamma$%
\begin{equation}
\mathcal{L}=-\frac{1}{2}\left( 1+\gamma m^{2}\right) \tilde{\varphi}\square%
\tilde{\varphi}-\frac{1}{2}m^{2}\tilde{\varphi}^{2}-\frac{\alpha}{4}\tilde{%
\varphi}^{4}-\frac{1}{2}\gamma\alpha\tilde{\varphi}^{3}\square \tilde{\varphi%
}.
\end{equation}
We can rewrite this expression in terms of the original field $\varphi$
perturbatively, to first order in $\gamma$, by using equation (\ref{EMB})%
\begin{equation}
\tilde{\varphi}\simeq\left( \varphi-\frac{1}{2}\gamma\square\varphi\right)
=\varphi+\frac{1}{2}\gamma\left( m^{2}\varphi+\alpha\varphi^{3}\right) .
\end{equation}
In this way we get the following Lagrangian density, to first order in $%
\gamma$%
\begin{align}
\mathcal{\tilde{L}}_{G}=-\left(\frac{1}{2}+\gamma m^{2}\right) &\left( %
\varphi\square\varphi+m^{2}\left( 1-\gamma m^{2}\right)
\varphi^{2}\right . \notag\\
&\left . +\frac{\alpha}{2}\left( 1+2\gamma m^{2}\right)
\varphi^{4}+2\gamma\alpha\varphi^{3}\square\varphi+%
\alpha^{2}\gamma\varphi ^{6}\right) .  \label{LEQMOT}
\end{align}
This Lagrangian density is identical to (\ref{LAGBGAMA1})\ which also
reproduces the one given by the double zero method. Summarizing, to first
order in\ $\gamma$ we have%
\begin{equation}
\mathcal{L}_{D}=\mathcal{L}_{DZ}=\mathcal{\tilde{L}}_{G}.
\end{equation}

\section{The two-particle scattering in the Bernard-Duncan scalar field with
a $\protect\varphi^{4}$ interaction}

In this section we discuss the dispersion of two scalar particles, with a
dynamics described by the Lagrangian density (\ref{BL}), which contains the
interaction term%
\begin{equation}
\mathcal{L}_{int}=-\frac{\alpha}{4}\varphi^{4}.
\end{equation}
To test the perturbative Hamiltonian constraint approach we analyze this
quantum process following two different approaches. In the first place we
directly compute the scattering amplitude at first order in $\gamma$ using
the results of the perturbative Hamiltonian constraint method. After this we
obtain the expression for the scattering amplitude exact in $\gamma$ from
the Lagrangian (\ref{BL}), from which we derive the corresponding amplitude
at first order in $\gamma$. We assume that the quantum Hamiltonians are
defined with the normal order product, so that the tadpole diagrams are not
considered.

It is interesting to study the dispersion of two scalar particles at first
order in $\gamma$, which in the effective theory involves not only
corrections to the original vertices, but also the new derivative vertex. To
proceed with the first calculation we consider this process using the
Feynman rules derived from the effective Lagrangian (\ref{LAGBGAMA1}). It
corresponds to the effective Hamiltonian (\ref{HAMB2}), obtained using the
perturbative Hamiltonian constraints approach, for which we have the
following algebra for the creation and annihilation operators%
\begin{equation}
\left[ a_{p},a_{p^{\prime}}^{\dagger}\right] =\left( 1-2\gamma m^{2}\right)
\delta\left( p-p^{\prime}\right) ,
\end{equation}
such that the normalized two $a$-particle $in$ and $out$ states are%
\begin{align}
\left\vert \Phi_{in}\right\rangle & =\frac{1}{\left( 1-2\gamma m^{2}\right) }%
a_{p_{1}}^{\dagger}a_{p_{2}}^{\dagger}\left\vert 0\right\rangle ,  \label{IN}
\\
\left\langle \Phi_{out}\right\vert & =\left\langle 0\right\vert
a_{p_{1}^{\prime}}a_{p_{2}^{\prime}}\frac{1}{\left( 1-2\gamma m^{2}\right) }.
\label{OUT}
\end{align}

\begin{figure}[pb]
\centerline{\psfig{file=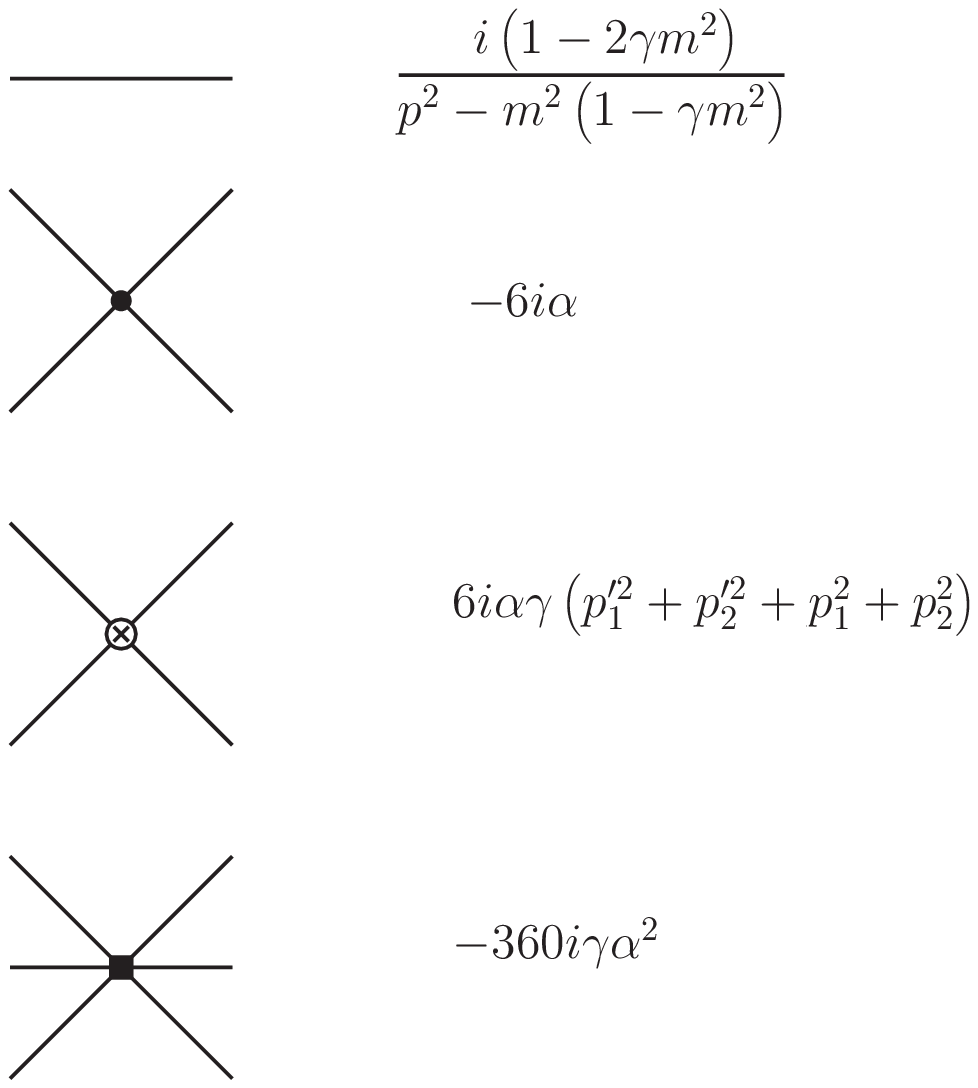,height=8cm}}
\vspace*{8pt}
\caption{Feynman rules for the effective scalar theory. Note that now we
have two additional vertices, a ${\protect\varphi^4}$ derivative vertex and
a ${\protect\varphi^6}$ vertex.}
\label{fig1}
\end{figure}

According to this, the first order contributions to the scattering amplitude
(see Fig. \ref{fig1}), are the ones given by the $\alpha^{4}$ vertex%
\begin{equation}
S_{fi}^{\left( 1\right) }=-\frac{3i\left( 2\pi\right) ^{4}\alpha}{2\sqrt{%
\omega_{p_{1}}\omega_{p_{2}}\omega_{p_{1}^{\prime}}\omega _{p_{2}^{\prime}}}}%
\delta^{4}\left( p_{1}^{\prime}+p_{2}^{\prime}-p_{1}-p_{2}\right) +O\left(
\gamma^{2}\right) ,
\end{equation}
and the derivative vertex%
\begin{equation}
S_{fi}^{\left( 1\right) }=\frac{6i\left( 2\pi\right) ^{4}\alpha\gamma m^{2}}{%
\sqrt{\omega_{p_{1}}\omega_{p_{2}}\omega_{p_{1}^{\prime}}\omega
_{p_{2}^{\prime}}}}\delta^{4}\left(
p_{1}^{\prime}+p_{2}^{\prime}-p_{1}-p_{2}\right) ,
\end{equation}
such that the total first order contribution results%
\begin{equation}
S_{fi}^{\left( 1\right) }=-\frac{3i\left( 2\pi\right) ^{4}\alpha\left(
1-4\gamma m^{2}\right) }{2\sqrt{\omega_{p_{1}}\omega_{p_{2}}\omega
_{p_{1}^{\prime}}\omega_{p_{2}^{\prime}}}}\delta^{4}\left( p_{1}^{\prime
}+p_{2}^{\prime}-p_{1}-p_{2}\right) .  \label{S1}
\end{equation}

\begin{figure}[pb]
\centerline{\psfig{file=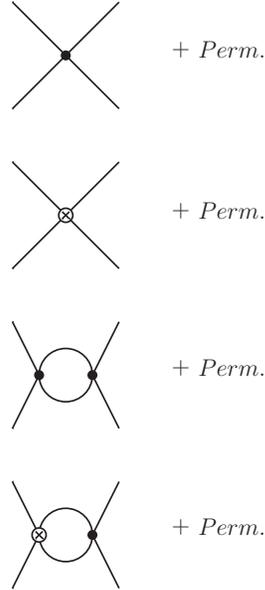,height=8cm}}
\vspace*{8pt}
\caption{Feynman diagrams contributing to the two-particle scattering in
the effective scalar theory, at first order in ${\protect\gamma}$ and one
loop approximation.}
\label{fig2}
\end{figure}

At second order we also have two contributions (see Fig. \ref{fig2}). One of them
corresponds to one loop with two $\alpha^{4}$ type vertices%
\begin{align}
S_{fi}^{\left( 2\right) } & =-\frac{9\alpha^{2}\left( 1-4\gamma m^{2}\right)
}{8\sqrt{\omega_{p_{1}}\omega_{p_{2}}\omega_{p_{1}^{\prime}}\omega_{p_{2}^{%
\prime}}}}\left( F\left( p_{1}+p_{2}\right) +F\left(
p_{1}^{\prime}-p_{1}\right) +F\left( p_{2}^{\prime}-p_{1}\right) \right)\notag\\
&\hskip6cm\times
\delta^{4}\left( p_{1}^{\prime}+p_{2}^{\prime}-p_{1}-p_{2}\right) , \\
F\left( p\right) & =\int d^{4}k\frac{1}{k^{2}-m^{2}\left( 1-\gamma
m^{2}\right) }\frac{1}{\left( p-k\right) ^{2}-m^{2}\left( 1-\gamma
m^{2}\right) },
\end{align}
and the other to one $\alpha^{4}$ type vertex and one derivative vertex%
\begin{align}
S_{fi}^{\left( 2\right) } & =-\frac{9\gamma\alpha^{2}}{2\sqrt{\omega
_{p_{1}}\omega_{p_{2}}\omega_{p_{1}^{\prime}}\omega_{p_{2}^{\prime}}}}\left(
G\left( p_{1}+p_{2}\right) +G\left( p_{1}^{\prime}-p_{1}\right) +G\left(
p_{2}^{\prime}-p_{1}\right) \right)\notag\\
&\hskip6cm\times \delta^{4}\left(
p_{1}^{\prime}+p_{2}^{\prime}-p_{1}-p_{2}\right) , \\
G\left( p\right) & \simeq\int d^{4}k\frac{-\left( 2m^{2}+k^{2}+\left(
p-k\right) ^{2}\right) }{k^{2}-m^{2}\left( 1-\gamma m^{2}\right) }\frac {1}{%
\left( p-k\right) ^{2}-m^{2}\left( 1-\gamma m^{2}\right) }.
\end{align}
The complete expression for the scattering amplitude in this approximation is%
\begin{align}
S_{fi}^{\left( 2\right) } & =-\frac{9\alpha^{2}\left( 1-4\gamma m^{2}\right)
}{2\sqrt{\omega_{p_{1}}\omega_{p_{2}}\omega_{p_{1}^{\prime}}\omega_{p_{2}^{%
\prime}}}}\delta^{4}\left( p_{1}^{\prime}+p_{2}^{\prime}-p_{1}-p_{2}\right)\notag\\
&\times\left( \left( F+\gamma G\right) _{\left( p_{1}+p_{2}\right)
}+\left( F+\gamma G\right) _{\left( p_{1}^{\prime }-p_{1}\right) }+\left(
F+\gamma G\right) _{\left( p_{2}^{\prime}-p_{1}\right) }\right)
 ,
\label{S2} \\
\left( F+\gamma G\right) _{\left( p\right) } & =\int d^{4}k\frac{1}{%
k^{2}-m^{2}\left( 1-\gamma m^{2}\right) }\frac{1-\left( 2m^{2}+k^{2}+\left(
p-k\right) ^{2}\right) }{\left( p-k\right) ^{2}-m^{2}\left( 1-\gamma
m^{2}\right) }.
\end{align}
Note that $F_{b}$ contains a quadratic divergence and a logarithmic
divergence. One should regularize this quantity in order to renormalize the
theory. However, the point we are interested in is comparing the results for
the scattering amplitude using two different methods, computing in the
effective Hamiltonian theory constructed using perturbative Hamiltonian
constraints in the usual fashion and computing in the exact HOTD theory, but
approximating the propagator and restricting the initial and final states.
For this reason we do not make explicit here the renormalization issues.

We now consider the same process, but in the framework of the exact
Lagrangian (\ref{BL}). To compare with the results from the perturbative
Hamiltonian constraints approach, in terms of the Fourier decomposition (\ref%
{PHIDESCOMP}), we consider only the asymptotic states corresponding to
physical, $a$-type, particles, not including negative norm ghost
contributions in the initial and final configurations. Thus, according to (%
\ref{ns}), the $in$ and $out$ states are respectively%
\begin{align}
\left\vert \Phi_{in}\right\rangle & =\sqrt{1+4\gamma m^{2}}%
a_{p_{1}}^{\dagger}a_{p_{2}}^{\dagger}\left\vert 0\right\rangle , \\
\left\langle \Phi_{out}\right\vert & =\left\langle 0\right\vert
a_{p_{1}^{\prime}}a_{p_{2}^{\prime}}\sqrt{1+4\gamma m^{2}}.
\end{align}

\begin{figure}[pb]
\centerline{\psfig{file=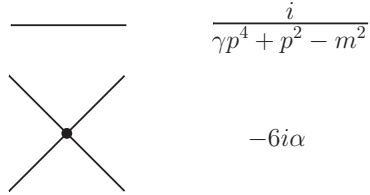,height=3cm}}
\vspace*{8pt}
\caption{Feynman rules for the HOTD Bernard-Duncan theory.}
\label{fig3}
\end{figure}

\begin{figure}[pb]
\centerline{\psfig{file=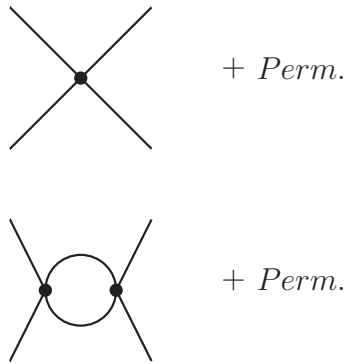,height=5cm}}
\vspace*{8pt}
\caption{Feynman diagrams contributing to the two-particle scattering in
the HOTD Bernard-Duncan theory, to first order in ${\protect\gamma}$ and one
loop approximation.}
\label{fig4}
\end{figure}

Hence, after adding all contributions up to first order in $\gamma$ see
Figs. (\ref{fig3}-\ref{fig4}), the scattering amplitude at first and second order in $\alpha$
are respectively%
\begin{align}
S_{fi}^{\left( 1\right) } & =\frac{-\alpha}{4}\int d^{4}x\left\langle
0\right\vert a_{p_{1}^{\prime}}a_{p_{2}^{\prime}}\text{ }\mathbf{:}\varphi
^{4}\left( x\right) \mathbf{:}a_{p_{1}}^{\dagger}a_{p_{2}}^{\dagger
}\left\vert 0\right\rangle \notag\\
&=-\frac{3i\left( 2\pi\right) ^{4}\alpha\left(
1-4\gamma m^{2}\right) }{2\sqrt{\omega_{p_{1}}\omega_{p_{2}}\omega
_{p_{1}^{\prime}}\omega_{p_{2}^{\prime}}}}\delta^{4}\left( p_{1}^{\prime
}+p_{2}^{\prime}-p_{1}-p_{2}\right) , \\
S_{fi}^{\left( 2\right) } & =\frac{1}{2!}\left( \frac{-i\alpha}{4}\right)
^{2}\int d^{4}x_{2}\int d^{4}x_{1}\left\langle 0\right\vert
a_{p_{1}^{\prime}}a_{p_{2}^{\prime}}\text{ }T\left( \mathbf{:}%
\varphi^{4}\left( x_{1}\right) \mathbf{:}\text{ }\mathbf{:}\varphi^{4}\left(
x_{2}\right) \mathbf{:}\text{ }\right)
a_{p_{1}}^{\dagger}a_{p_{2}}^{\dagger}\left\vert 0\right\rangle  \notag \\
& =-\frac{9\alpha^{2}\left( 1-4\gamma m^{2}\right) }{2\sqrt{%
\omega_{p_{1}}\omega_{p_{2}}\omega_{p_{1}^{\prime}}\omega_{p_{2}^{\prime}}}}%
\delta ^{4}\left( p_{1}^{\prime}+p_{2}^{\prime}-p_{1}-p_{2}\right) \notag\\
&\hskip4cm\times\left(
\tilde{F}\left( p_{1}+p_{2}\right) +\tilde{F}\left(
p_{1}^{\prime}-p_{1}\right) +\tilde{F}\left( p_{2}^{\prime}-p_{1}\right)
\right) ,
\end{align}
where%
\begin{equation}
\tilde{F}\left( p\right) =\int d^{4}k\frac{1}{\gamma k^{4}+k^{2}-m^{2}}\frac{%
1}{\gamma\left( k-p\right) ^{4}+\left( k-p\right) ^{2}-m^{2}}.  \label{FP}
\end{equation}
At first order in $\gamma$ they reduce to (\ref{S1}) and (\ref{S2}), with $%
\tilde{F}\left( p\right) =\left( F+\gamma G\right) _{\left( p\right) }+%
\mathcal{O}(\gamma^{2})$. Both computations, the one based on perturbative
Hamiltonians constraints and the one obtained from a perturbative expansion
of the propagator of the exact Lagrangian, where we have used a procedure
similar to the one applied by Weinberg in Ref. \refcite{WEINBERG}, yield the
same result at the order $\gamma$ considered. This is evidence of the
soundness of the effective Hamiltonian construction proposed here.

\section{Final remarks}

In this article we have presented an alternative method for constructing a
consistent effective Hamiltonian formalism for higher order Lagrangians,
which gives a correct low energy approximation. It includes higher energy
scale effects under the form of perturbative corrections in the framework of
a second order theory, so that it is free from the pathological
behavior characteristic of HOTD theories. This method results from
an application of the standard Dirac procedure to deal with constrained
systems
which, together with the constraints generated by the definition of
the momenta and their consistency conditions, incorporates a new
set of perturbative constraints generated by the inhomogeneous relations in
the perturbative parameter contained in the set of the original momenta
definitions and constraints. The addition of these new constraints
projects the dynamics of the original HOTD system into a
stable subspace consistent with the chosen order of approximation.
Our method is purely algebraic and does no require to solve any system of
differential equations. Also, it can be directly applied to gauge
systems without the necessity of initially fixing the gauge. Moreover, we conjecture
that the brackets obtained from this procedure in phase space, when rewritten
in the CV space via the corresponding equations of motion, would provide a solution
for the system of differential equations (\ref{SYSTEM}), which is the starting point of the
Eliezer-Woodard formulation \cite{EW}. The resulting  second order effective Lagrangians produce unitary
theories after quantization. The study of the possible breaking of Lorentz covariance induced by the perturbative
description is beyond the scope of the present work \cite{BD}.

At the classical level this Hamiltonian formalism yields canonical equations
of motion equivalent to the usual perturbative approximation for the
Lagrangian equations of motion of the exact HOTD theory. Furthermore, from
this canonical construction we can also derive a well behaved effective
Lagrangian formalism. At the quantum level we recover the results obtained
using the perturbative expression for the propagators, provided that the
asymptotic state space is restricted to the physical one, ruling out the
ghosts. The difference with this last approach is that now we have a well
defined canonical formalism, instead of a non-consistent one where ghosts
must be forbidden because asymptotic states and approximate propagators are
used. This new formulation provides a consistent effective theory,
which allows the implementation of all the usual manipulations for the
construction of a quantum field theory.

An important characteristic of our approach is that it makes no use at all
of the Lagrangian equations of motion or any Lagrangian constraints to
obtain the corresponding effective theory. This is a significant difference
with previous proposals, particularly the methods presented in Refs. \refcite{EW} and  \refcite%
{JLM}. Moreover, our formalism can deal with singular theories
in a straightforward way, since we only have to add the corresponding non
perturbative relations to the total set of constraints and work within the
Dirac framework in the usual way. Another characteristic is that the
procedure is not an iterative one. At the start we choose the order of the
approximation, which is defined by the chosen perturbative
constraints, and the Dirac algorithm converges directly to the
corresponding effective theory.

As working examples, we have applied the method to two paradigmatic models
in HOTD theories: the Pais-Uhlenbeck oscillator and the Bernard-Duncan
scalar field, in which we include a $\varphi^{4}$ interaction. The first
example clearly shows the main features of the perturbative Hamiltonian
constraint approach, and allows us to express the perturbative constraints
in a very simple and closed way, in terms of Fibonacci polynomials. In the
other example, closed expressions for the constraints are much more
difficult to write and for this reason we restricted the construction
only to first order in the perturbative parameter. In both cases
the canonical equations of motion are equivalent to the Lagrangian
perturbative ones.

From the Hamiltonian thus constructed we can obtain a second order effective
Lagrangian. The quantum theory can be obtained either from the
Hamiltonian formalism or from the effective Lagrangian approach,
via the Matthews' theorem which is well established for Lagrangians of the
form here obtained\cite{BD}. The calculation of the two-particle scattering
shows that the effective theory constructed on the basis of perturbative
Hamiltonian constraints gives the same results as the exact higher order
theory, provided that in this last theory the space of states is restricted
to the physical one, ruling out ghost states, and that the propagator
is considered in terms of a perturbative expansion. Similar results
are obtained in Ref. \refcite{WEINBERG}, although there a direct
substitution of leading order equations of motion was performed in the
Lagrangian to get the effective theory. As it has been clearly shown in Ref. %
\refcite{BARKEROCONELL}, this procedure can be considered to be
correct only when it effectively results in the addition of a double zero
term to the original Lagrangian, or when a suitable equivalent
derivative field transformation is found.

\section*{Acknowledgments}

RM and LFU would like to thanks useful discussions with J. D. Vergara. LFU
is partially supported by projects CONACYT \# 55310 and DGAPA-UNAM-IN111210. He also
acknowledges support from RED-FAE, CONACYT. SM and RM are supported by CONICET-Argentina.

\appendix

\section{}

In this Appendix we explore the connection between Lagrangian and
Hamiltonian perturbative constraints, in the context of a simple example of
HOTD theory. Let us consider the Lagrangian%
\begin{equation}
L_{g}=\frac{1}{2}\left( \dot{q}_{i}\right) ^{2}-\frac{1}{2}\omega ^{2}\left(
q_{i}\right) ^{2}+\frac{\gamma }{2}\epsilon _{ij}\dot{q}_{i}\ddot{q}_{j}. \qquad  i=1,2
\label{LAGEX}
\end{equation}%
For simplicity, we will construct the  Hamiltonian formalism only to  first order in $%
\gamma $. The exact equations of motion are%
\begin{equation}
\ddot{q}_{i}+\gamma \epsilon _{ij}q_{j}^{\left( 3\right) }+\omega ^{2}q_{i}=0
\end{equation}%
which, to first order in $\gamma $, reduce to%
\begin{equation}
\ddot{q}_{i}= \gamma \omega ^{2}\epsilon _{ij}\dot{q}_{j}-\omega ^{2}q_{i}
\label{EQ1GAMMA}
\end{equation}%
In this case, the approach of Jaen, Llosa and Molina \cite{JLM} and the one of Eliezer and Woodard \cite{EW}
are equivalent, as stated in Ref. \refcite{EW1}. For this reason we will consider only
the second one, which is more adequate to establish the comparison. Following Ref. \refcite{EW} we calculate the Noether energy
\begin{equation}
E(q,{\dot q})=\frac{1}{2} \dot{q}_{i}^{2}+\frac{1}{2}\omega ^{2}
q_{i}^{2}+\gamma \epsilon _{ij}\dot{q}_{i}\ddot{q}_{j},
\end{equation}%
which is taken as the time evolution generator in the coordinate-velocity (CV) space. In principle $E(q,{\dot q})$
should be a  projection on the CV space of the
Hamiltonian which lives in a  not yet known phase space.
To first order in $\gamma$ the above equation  yields%
\begin{equation}
E_{1}=\frac{1}{2} \dot{q}_{i}^{2}+\frac{1}{2}\omega ^{2}
q_{i}^{2}+\gamma \omega ^{2}\epsilon _{ij}q_{i}\dot{q}_{j}.
\label{E1}
\end{equation}%

The method  further  assumes the existence of a fundamental
bracket $\left\{ q_{i}, {q}_{j}\right\} $ in terms of which we can describe the temporal evolution, such that%
\begin{align}
\dot{q}_{i}& =\left\{ q_{i},E_{1}\right\} =\left\{ q_{i},q_{j}\right\} \frac{%
\partial E_{1}}{\partial q_{j}}+\left\{ q_{i},\dot{q}_{j}\right\} \frac{%
\partial E_{1}}{\partial \dot{q}_{j}}, \\
\ddot{q}_{i}& =\left\{ \dot{q}_{i},E_{1}\right\} =\left\{ \dot{q}%
_{i},q_{j}\right\} \frac{\partial E_{1}}{\partial q_{j}}+\left\{ \dot{q}_{i},%
\dot{q}_{j}\right\} \frac{\partial E_{1}}{\partial \dot{q}_{j}}.
\end{align}%
These brackets must also satisfy the consistency conditions%
\begin{align}
\left\{ \dot{q}_{i},q_{j}\right\} +\left\{ q_{i},\dot{q}_{j}\right\} & =%
\frac{d}{dt}\left\{ q_{i},q_{j}\right\},  \\
\left\{ \dot{q}_{i},\dot{q}_{j}\right\} & =\left( \frac{1}{2}\frac{d^{2}}{%
dt^{2}}+2\omega ^{2}\right) \left\{ q_{i},q_{j}\right\}
\end{align}%
and thus we obtain the following system of differential equations defining the  basic objects in our
example
\begin{align}
\dot{q}_{i}=&\left( \omega ^{2}q_{j}+\gamma \omega ^{2}\epsilon _{jk}\dot{q}%
_{k}\right) \left\{ q_{i},q_{j}\right\} +\left( \dot{q}_{j}-\gamma \omega
^{2}\epsilon _{jk}q_{k}\right) \left\{ q_{i},\dot{q}_{j}\right\}, \nonumber \\
\ddot{q}_{i} =&\left( \omega ^{2}q_{j}+\gamma \omega ^{2}\epsilon _{jk}\dot{q%
}_{k}\right) \left( \frac{d}{dt}\left\{ q_{i},q_{j}\right\} -\left\{ q_{i},%
\dot{q}_{j}\right\} \right) \nonumber\\
&+\left( \dot{q}_{j}-\gamma \omega ^{2}\epsilon
_{jk}q_{k}\right) \left( \frac{1}{2}\frac{d^{2}}{dt^{2}}+2\omega ^{2}\right)
\left\{ q_{i},q_{j}\right\},
\label{SYSTEM}
\end{align}%
where $\ddot{q}_{i}$ is given by Eq. (\ref{EQ1GAMMA}). Following this approach, it is
necessary to find the adequate solution to the above system,  which in this case must be analytical in $q_{i}$
and $\dot{q}_{i}$. Next, from this solution  the
relation between canonical momenta, coordinates and velocities  must be inferred. Once this is
achieved, the canonical formalism to
first order in $\gamma $ is constructed by  projecting the energy $E_{1}$ together with the brackets among coordinates
and velocities in the phase space just defined.  Additional simplification of the system (\ref{SYSTEM}) can be achieved by making the
following ansatz
\begin{equation}
\left\{ q_{i},q_{j}\right\}=\gamma A_{ij}, \qquad \left\{ q_{i},\dot{q}_{j}\right\}=\delta_{ij}+\gamma B_{ij},
\end{equation}
to first order in $\gamma$. Even with the above simplification, some guess work has to be done in order to solve the Eqs. (\ref{SYSTEM}). It is clear that
the complexity of the basic equations (\ref{SYSTEM}) will rapidly increase either when higher order approximations are considered or when more
complicated systems are studied. We consider this as a shortcoming of the method proposed in Ref. \refcite{EW}.

 For the  purpose of  comparing the method based on Lagrangian
constraints, in the Eliezer and Woodard incarnation, with the one based on the Dirac approach, the expressions already
obtained are enough. Now we will deal with the problem using the Dirac method
augmented with perturbative Hamiltonian constraints. To apply the procedure
we first rewrite the Lagrangian (\ref{LAGEX}) in first order form by introducing the coordinates $z_i$ via the auxiliary coordinates $\lambda_i$
\begin{equation}
L_{g}=\frac{1}{2}\left( z_{i}\right) ^{2}-\frac{1}{2}\omega ^{2}\left(
q_{i}\right) ^{2}+\frac{\gamma }{2}\epsilon _{ij}z_{i}\dot{z}_{j}+\lambda
_{i}\left( z_{i}-\dot{q}_{i}\right).
\end{equation}%
The canonical momenta are%
\begin{equation}
p_{q_{i}}=-\lambda _{i},\ \ \ \ \ p_{\lambda _{i}}=0,\ \ \ \ \pi _{z_{i}}=%
\frac{\gamma }{2}\epsilon _{ji}z_{j}.
\end{equation}%
So we have six exact primary constraints, plus two pertubative ones%
\begin{equation}
\gamma \pi_{z_{i}}=0.
\end{equation}%
Following with the Dirac method  we  demand the consistency of the
constraints under time evolution and we finally arrive to a Dirac Hamiltonian. Once the second class
constraints are imposed as strong relations it reduces to%
\begin{equation}
H_{C}=\frac{1}{2} p_{i}^{2}+\frac{1}{2}\omega ^{2}
q_{i}^{2},
\end{equation}%
with the Dirac brackets%
\begin{equation}
\left\{ q_{i},p_{j}\right\}_D =\delta _{ij}, \qquad \left\{
p_{i},p_{j}\right\}_D =0, \qquad \left\{ q_{i},q_{j}\right\}_D =\gamma \epsilon.
_{ij}
\label{DBEX}
\end{equation}%
To project this formalism in the CV space and compare with the
Eliezer-Woodard approach, we perform the transformation $\left( q,\dot{q}\right)
\rightarrow \left( q,p\right)$ which is given by the canonical equations of motion%
\begin{equation}
\dot{q}_{i}=\left\{ q_{i},H_{C}\right\} =p_{i}+\gamma \omega ^{2}\epsilon
_{ij}q_{j}
\label{TRANSF}
\end{equation}%
Projecting $H_{C}$ we obtain%
\begin{equation}
H_{C}=\frac{1}{2} \dot{q}_{i}^{2}+\frac{1}{2}\omega ^{2}
q_{i}^{2}+\gamma \omega ^{2}\epsilon _{ij}q_{i}\dot{q}_{j},
\end{equation}%
which coincides with the Noether  energy restricted to the CV subspace (\ref{E1}).
From the Dirac brackets (\ref{DBEX}) and
the transformation (\ref{TRANSF}) we can compute
\begin{align}
\left\{ q_{i},\dot{q}_{j}\right\} & =\left\{ q_{i},p_{j}+\gamma \omega
^{2}\epsilon _{jk}q_{k}\right\} =\delta _{ij}, \\
\left\{ \dot{q}_{i},\dot{q}_{j}\right\} & =\left\{ p_{i}+\gamma \omega
^{2}\epsilon _{im}q_{m},p_{j}+\gamma \omega ^{2}\epsilon _{jk}q_{k}\right\}
=2\gamma \omega ^{2}\epsilon _{ij}.
\end{align}%
It is straightforward to verify that this set of brackets is indeed a solution of
the set of diferential equations (\ref{SYSTEM}).

\end{document}